\documentclass[fleqn]{annalen}
\usepackage{epsf}
\def\figuresize{8.4cm}

\begin{document}

%%%%%%%%%%%%%%%%%%%%%%%%%%%%%%%%%%%%%%%%%%%%%%%%%%%%%%%%%%%%%%%%%%%%%%%%%%%%%%
%%%%%%%% the following newcommands will be completed by the publisher %%%%%%%%
%%%%%%%%%%%%%%%%%%%%%%%%%%%%%%%%%%%%%%%%%%%%%%%%%%%%%%%%%%%%%%%%%%%%%%%%%%%%%%
\newcommand{\volume}{8}              %sets current volume,
\newcommand{\xyear}{1999}            %sets year in header
\newcommand{\issue}{Spec. Issue, SI-3 -- SI-9}               %sets current issue,
\newcommand{\recdate}{29 July 1999}  %sets received date,
\newcommand{\revdate}{dd.mm.yyyy}    %sets revised date,
\newcommand{\revnum}{0}              %number of revisions,
\newcommand{\accdate}{dd.mm.yyyy}    %sets accepted date,
\newcommand{\coeditor}{ue}           %sets (co)editor,
\newcommand{\firstpage}{61}         %first page number,
\newcommand{\lastpage}{64}          %last page number,
\setcounter{page}{\firstpage}        %sets page counter to first page number
%%%%%%%%%%%%%%%%%%%%%%%%%%%%%%%%%%%%%%%%%%%%%%%%%%%%%%%%%%%%%%%%%%%%%%%%%%%%%%

\newcommand{\keywords}{localization, interactions, quantum Coulomb glass}

\newcommand{\PACS}{75.20.Hr, 75.10.Jm, 75.30.Kz}

%%%%%%%%%%%%%%%%%%%%%%%%%%%%%%%%%%%%%%%%%%%%%%%%%%%%%%%%%%%%%%%%%%%%%%%%%%%%%%
%% please enter (First) Author (et al.) and short version of the title here %%
%%%%%%%%%%%% must not exceed 80 characters in length together %%%%%%%%%%%%%%%%
%%%%%%%%%%%%%%%%%%%%%%%%%%%%%%%%%%%%%%%%%%%%%%%%%%%%%%%%%%%%%%%%%%%%%%%%%%%%%%

\newcommand{\shorttitle}{
F.\ Epperlein et al., Crossover from interaction-induced localization to
delocalization}

%%%%%%%%%%%%%%%%%%%%%%%%%%%%%%%%%%%%%%%%%%%%%%%%%%%%%%%%%%%%%%%%%%%%%%%%%%%%%%

\title{Crossover from interaction induced localization to delocalization in
disordered electron systems}

%%%%%%%%%%%%%%%%%%%%%%%%%%%%%%%%%%%%%%%%%%%%%%%%%%%%%%%%%%%%%%%%%%%%%%%%%%%%%%

\author{Frank Epperlein$^{1}$, Thomas Vojta$^{1,2}$, and
    Michael Schreiber$^{1}$}

%%%%%%%%%%%%%%%%%%%%%%%%%%%%%%%%%%%%%%%%%%%%%%%%%%%%%%%%%%%%%%%%%%%%%%%%%%%%%%

\newcommand{\address}
{$^1$Institut f{\"u}r Physik, TU Chemnitz, D-09107 Chemnitz, FRG\\
 $^2$Materials Science Institute,
   University of Oregon, Eugene, OR 97403, USA}

%%%%%%%%%%%%%%%%%%%%%%%%%%%%%%%%%%%%%%%%%%%%%%%%%%%%%%%%%%%%%%%%%%%%%%%%%%%%%%

\newcommand{\email}{\tt vojta@physik.tu-chemnitz.de}

\maketitle

%%%%%%%%%%%%%%%%%%%%%%%%%%%%%%%%%%%%%%%%%%%%%%%%%%%%%%%%%%%%%%%%%%%%%%%%%%%%%

% macros
\def\b{\bibitem}

%%%%%%%%%%%%%%%%%%%%%%%%%%%%%%%%%%%%%%%%%%%%%%%%%%%%%%%%%%%%%%%%%%%%%%%%%%%%%%

\begin{abstract}
We numerically investigate the transport properties of interacting spinless
electrons in disordered systems. We use an efficient method which is based
on the diagonalization of the Hamiltonian in the subspace of the
many-particle Hilbert space which is spanned by the low-energy Slater
states. Low-energy properties can be calculated with an accuracy comparable
to that of exact diagonalization but for larger system sizes. The method
works well in the entire parameter space, and it can handle long-range as
well as short-range interactions.
Using this method we calculate the combined effect of disorder and
interactions on the Kubo-Greenwood conductance and on the sensitivity of the
ground state energy to a twist in the boundary conditions. We find that the
influence of the interactions on the transport properties is opposite for
large and small disorder. In the strongly localized regime (small kinetic
energy, large disorder) interactions increase the transport whereas for weak
disorder (large kinetic energy) interactions decrease the transport.
\end{abstract}

%%%%%%%%%%%%%%%%%%%%%%%%%%%%%%%%%%%%%%%%%%%%%%%%%%%%%%%%%%%%%%%%%%%%%%%%%%%%%
\section{Introduction}
%%%%%%%%%%%%%%%%%%%%%%%%%%%%%%%%%%%%%%%%%%%%%%%%%%%%%%%%%%%%%%%%%%%%%%%%%%%%%
\label{sec:I}
The influence of electron-electron interactions on Anderson localization
has reattracted a lot of attention, in particular because recent
experimental results show a metal-insulator transition (MIT) in the
two-dimensional electron gas in Si-MOSFETs \cite{2dMIT}. This is in
conflict with the theory of Anderson localization for non-interacting
electrons. Since the electron density in the Si-MOSFETs is very low which
makes the electron-electron interaction particularly important, it is
generally assumed that some type of interaction effect is responsible for
the MIT. A complete understanding has, however, not yet been obtained.
There have been attempts to explain the experiments based on the
perturbative renormalization group \cite{runaway}, non-perturbative effects
\cite{nonperturb}, or the transition being a superconductor-insulator
transition rather than a MIT \cite{SIT}.

We have investigated the influence of disorder and interactions on
the transport of spinless electrons numerically. Within an
effective single-particle theory at the Hartree-Fock level
interactions always have a localizing effect \cite{Hf}. In order to go
beyond the Hartree-Fock approximation we have developed \cite{Dointer,HFD}
an efficient method, the Hartree-Fock based diagonalization (HFD)
which is related to the quantum-chemical configuration interaction
approach. We have used this method to study the influence of
interactions on the conductance in one
\cite{Transportin}, two \cite{Dointer}, and three \cite{pils98} dimensions.
We found a delocalizing tendency of the interactions for strong disorder
but a localizing one for weak disorder. Similar results have been obtained
by means of the density-matrix renormalization group \cite{schmitt} in one
dimension and exact diagonalization in two dimensions \cite{benenti}.

In this paper we compare the results for the Kubo-Greenwood conductance to
those for the phase sensitivity, i.e., the reaction of the ground state
energy to a twist in the boundary conditions. We only summarize the basic
findings, the details will be published elsewhere.

%%%%%%%%%%%%%%%%%%%%%%%%%%%%%%%%%%%%%%%%%%%%%%%%%%%%%%%%%%%%%%%%%%%%%%%%%%%%%
\section{Quantum Coulomb glass and Hartree-Fock based diagonalization}
%%%%%%%%%%%%%%%%%%%%%%%%%%%%%%%%%%%%%%%%%%%%%%%%%%%%%%%%%%%%%%%%%%%%%%%%%%%%%
\label{sec:II}

The generic model for spinless interacting disordered electrons is the
quantum Coulomb glass \cite{Hf,TPE96}. It is defined on a regular
hypercubic lattice with $g=L^d$ ($d$ is the spatial dimensionality) sites
occupied by $N=K g$ electrons ($0\!<\!K\!<\!1$). To ensure charge
neutrality each lattice site carries a compensating positive charge of
$Ke$. The Hamiltonian is given by
\begin{equation}
H =  -t  \sum_{\langle ij\rangle} (c_i^\dagger c_j + c_j^\dagger c_i) +
       \sum_i \varphi_i  n_i + \frac{1}{2}\sum_{i\not=j}(n_i-K)(n_j-K)U_{ij}
\label{eq:Hamiltonian}
\end{equation}
where $c_i^\dagger$ and $c_i$ are the electron creation and annihilation
operators at site $i$, respectively, and $\langle ij \rangle$ denotes all
pairs of nearest-neighbor sites. $t$ is the strength of the hopping term,
i.e., the kinetic energy, and $n_i$ is the occupation number of site $i$.
We parametrize the Coulomb interaction $U_{ij} = e^2/r_{ij}$ by its value
$U$ between nearest-neighbor sites. The random potential values
$\varphi_i$ are chosen independently from a box distribution of width $2
W_0$ and zero mean. The boundary conditions are periodic with an additional
Bloch phase in one of the dimensions. The Coulomb interaction is treated in
the minimum image convention.

A numerically exact solution of the quantum Coulomb glass requires the
diagonalization of a matrix whose dimension increases exponentially with
system size. This severely limits the possible sample sizes. In order to
overcome this problem we have developed the HFD method. The basic idea is
to work in a truncated Hilbert space consisting of the corresponding
Hartree-Fock (Slater) ground state and the low-lying excited Slater states.
For each disorder configuration three steps have to be performed: (i) find
the Hartree-Fock solution of the problem, (ii) determine the $B$ Slater
states with the lowest energies, and (iii) calculate and diagonalize the
Hamilton matrix in the subspace spanned by these states. The number $B$ of
new basis states determines the quality of the approximation, reasonable
values have to be found empirically.

%%%%%%%%%%%%%%%%%%%%%%%%%%%%%%%%%%%%%%%%%%%%%%%%%%%%%%%%%%%%%%%%%%%%%%%%%%%%%
\section{Kubo-Greenwood conductance and phase sensitivity}
%%%%%%%%%%%%%%%%%%%%%%%%%%%%%%%%%%%%%%%%%%%%%%%%%%%%%%%%%%%%%%%%%%%%%%%%%%%%%
\label{sec:III}

\begin{figure}
  \epsfxsize=\figuresize
  \centerline{\epsffile{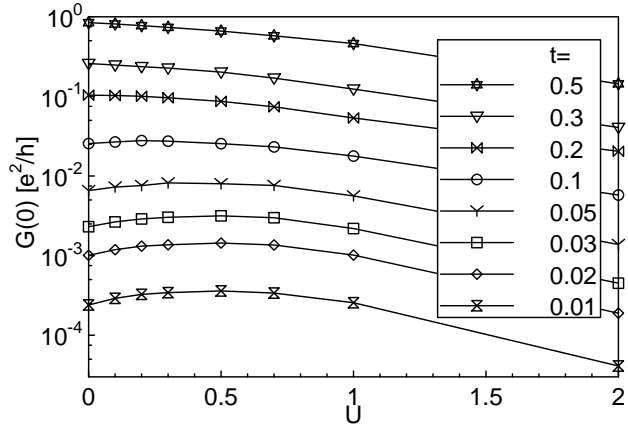}}
  \caption{Logarithmically averaged (400 samples) d.c. conductance $G(0)$
              for systems of
              $4 \times 4$ lattice sites and 8 electrons for different
              $U$ and $t$. The disorder  strength is fixed to $W_0=1$,
              the broadening is $\gamma=0.05$, and
              the HFD basis size is $B=300$.}
  \label{Fig:conductance}
\end{figure}

In order to characterize the transport behavior we have studied two
quantities, the Kubo-Greenwood conductance and the phase sensitivity of the
ground-state energy. The conductance \cite{K57,G58} can be
obtained from linear-response theory. It is essentially given by the
current-current correlation function of the ground state. For an isolated
finite system the conductance as a function of frequency consists of a
finite number of $\delta$ peaks at the excitation energies of the system.
In order to extrapolate to the d.c.\ conductance at zero frequency an
inhomogeneous broadening $\gamma$ is assumed in our calculations which
mimics the coupling to leads and contacts. This broadening introduces
an additional empirical parameter. We have calculated the conductance
for systems with up to 100 lattice sites. In Fig.\
\ref{Fig:conductance} we show a typical result (for a comparatively small
system which also allows the calculation of the phase sensitivity).
For weak disorder (large kinetic energy $t$) the interactions always reduce
the conductance while for strong disorder (small $t$) moderate
interactions significantly increase the conductance. Sufficiently strong
interactions always strongly suppress the conductance. This is the
precursor of a Wigner crystal or Wigner glass. The behavior of the conductance
can be attributed to the competition of two effects: First, the interactions destroy
the phase of the electrons and thus the interference necessary for
localization. This is particularly effective if the localization length is
small to begin with. Second, the interactions introduce an additional source
of randomness which tends to increase the localization.

\begin{figure}
  \epsfxsize=\figuresize
  \centerline{\epsffile{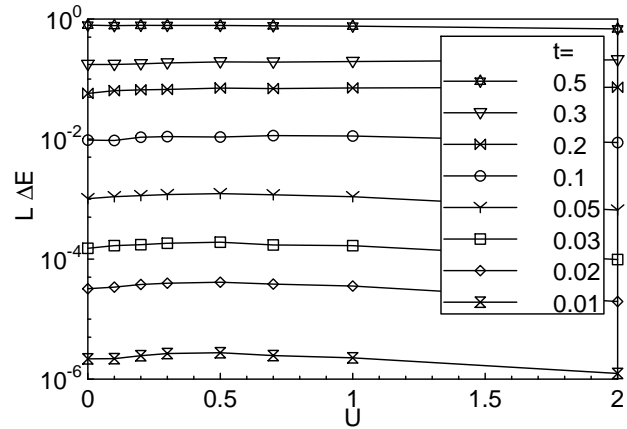}}
  \caption{Logarithmically averaged (400 samples) phase sensitivity
              $L_x ~\Delta E$. The parameters are as in Fig.
              \protect{\ref{Fig:conductance}}.
              The basis size within the HFD method was $B=500$.}
  \label{Fig:sensitivity}
\end{figure}

It is well known \cite{montambaux} that in disordered systems there is a
close relation between the charge
 stiffness (i.e., the 2nd derivative of the
ground state energy with respect to a magnetic flux) and the conductance. A
numerically simpler but related quantity is the phase sensitivity of the ground
state. It is given by the difference between the ground-state energies for
periodic and antiperiodic boundary conditions \cite{crossings}. In Fig.\
\ref{Fig:sensitivity} we show numerical results for the phase sensitivity
of the same systems as considered in Fig. \ref{Fig:conductance}.
For strong disorder (small $t$) the phase sensitivity shows the same
qualitative behavior as the conductance. However, the
interaction-induced enhancement of the phase sensitivity for moderate
interactions is weaker than the enhancement of the conductance. For small
disorder (large $t$) a crossover to a localizing effect of the interactions
starts, but to really reach the localizing regime seems to require still
larger $t$. The differences between the results for the Kubo-Greenwood
conductance and the phase sensitivity are not completely understood so far.
However, a likely reason are the ambiguities in describing the contacts and
in taking the zero-frequency limit.

In summary, we have studied the influence of electron-electron interactions
on Anderson localization for spinless electrons in two dimensions. For strong
disorder moderate interactions significantly enhance the transport.

%%%%%%%%%%%%%%%%%%%%%%%%%%%%%%%%%%%%%%%%%%%%%%%%%%%%%%%%%%%%%%%%%%%%%%%%%%
\vspace*{0.25cm} \baselineskip=10pt{\small \noindent
This work was supported in part by the DFG (SFB 393/C2). T.V. thanks the
Aspen Center for Physics for hospitality during the completion of this
paper.}
%%%%%%%%%%%%%%%%%%%%%%%%%%%%%%%%%%%%%%%%%%%%%%%%%%%%%%%%%%%%%%%%%%%%%%%%%%

\end{document}